\documentstyle[11pt,th-en-roman]{article}
\def\la{\leftarrow}
\def\ra{\rightarrow}

\def\bd{\noindent\bf}

\def\beginex#1{\trivlist \item[\hskip \labelsep{\bf #1}]}
\def\endex{\endtrivlist}

\newtheorem{Def}{Definition}
\newtheorem{Defs}[Def]{Definitions}
\newtheorem{Prop}{Proposition}
\newtheorem{Lem}{Lemma}

\begin{document}
\date{}
\title{\Large\bf Third Order Matching is Decidable}

\author{Gilles Dowek\\
        INRIA-Rocquencourt,\\
        B.P. 105, 78153 Le Chesnay Cedex, France\\
        Gilles.Dowek@inria.fr}
\maketitle

\thispagestyle{empty}

\subsection*{Abstract} 

{\it The {\em higher order matching} problem is the problem of determining 
whether a term is an instance of another in the simply typed 
$\lambda$-calculus, i.e. to solve the equation $a = b$ where $a$ and $b$ 
are simply typed $\lambda$-terms and $b$ is ground. The decidability of 
this problem is still open. We prove the decidability of the particular
case in which the variables occurring in the problem are at most third 
order.}

\section*{Introduction}

The {\em higher order matching} problem is the problem of determining whether
a term is an instance of another in the simply typed $\lambda$-calculus
i.e. to solve the equation $a = b$ where $a$ and $b$ are simply typed
$\lambda$-terms and $b$ is ground. 

Pattern matching algorithms are used to check if a proposition can be
deduced 
from another by elimination of universal quantifiers or by introduction of
existential quantifiers. In automated theorem proving, elimination of 
universal quantifiers and introduction of existential quantifiers are mixed
and full unification is required, but in proof-checking and semi-automated
theorem proving, these rules can be applied separately and thus pattern
matching can be used instead of unification.

Higher order matching is conjectured decidable in \cite{Huet76} and 
the problem is still open.
In \cite{Huet75} \cite{Huet76} \cite{HueLan} Huet has given a semi-decision 
algorithm and shown that in the particular case in which the variables 
occurring in the term $a$ are at most second order this algorithm terminates, 
and thus that second order matching is decidable. In \cite{Statman} Statman 
has reduced the conjecture to the $\lambda$-definability conjecture and in 
\cite{Wolfram} Wolfram has given an always terminating algorithm whose 
completeness is conjectured.

We prove in this paper that third order matching is decidable i.e. we give an 
algorithm that decides if a matching problem, in which all the variables are 
at most third order, has a solution. The main idea is that if the problem
$a = b$ has a solution then it also has a solution whose depth is bounded by 
some integer $s$ depending only on the problem $a = b$, so a simple 
enumeration
of the substitutions whose depth is bounded by $s$ gives a decision algorithm.
This result can also be used to bound the depth of the search tree in Huet's 
semi-decision algorithm and thus turn it into a always-terminating
decision algorithm.
It can also be used to design an algorithm which enumerates a complete
set of solutions to a third order matching problem and either terminates
if the problem has a finite complete set of solutions or keeps
enumerating solutions forever if it the problem admits no such set.
At last we discuss the problems that occur when we try to generalize the 
proof given here to higher order matching.

\section{Trees and Terms}

\subsection{Trees}

\begin{Defs} (Following \cite{Gorn})
An {\it occurrence} is a list of strictly positive integers
$\alpha = <s_{1}, ..., s_{n}>$. The number $n$ is called the 
{\it length} of the occurrence $\alpha$.
A {\it tree domain} $D$ is a non empty finite set of occurrences such that if 
$\alpha<n> \in D$ then $\alpha \in D$ and if also $n \neq  1$ then 
$\alpha<n-1> \in D$. A {\it tree} is a function from a tree domain $D$ to a 
set $L$, called the set of labels of the tree.

If $T$ is a tree and $D$ its domain, the occurrence $<~>$ is called the 
{\it root} of $T$ and the occurrence $\alpha<n>$ is called the {\it $n^{th}$
son} of the occurrence $\alpha$. The {\it number of sons} of an occurrence 
$\alpha$ is the greatest integer $n$ such that $\alpha<n> \in D$. A {\it leaf} 
is an occurrence that has no sons. 

Let $T$ be a tree and let $\alpha = <s_{1}, ..., s_{n}>$ be an occurrence in
this tree, the {\it path} of $\alpha$ is the set of occurrences 
$\{<s_{1}, ..., s_{p}>~|~p \leq n\}$. The number of elements of this path
is the length of $\alpha$ plus one.

The {\it depth} of the tree $T$ is the length of the longest occurrence in $D$.
This occurrence is, of course, a leaf. 

If $T$ is a tree of domain $D$ and $\alpha$ is an occurrence of $D$, the 
{\it subtree}
$T/\alpha$ is the tree $T'$ whose  domain is 
$D' = \{\beta~|~\alpha \beta \in D\}$ 
and such that 
$$T'(\beta) = T(\alpha \beta)$$
By an abuse of language, if $\alpha<n>$ is an occurrence of a tree $T$, the 
subtree $T/\alpha<n>$ is also called the {\it $n^{th}$ son} of the occurrence
$\alpha$. 

If $a$ is a label and $T_{1}, ..., T_{n}$ are trees (of domains 
$D_{1}, ..., D_{n}$) then the {\it tree of root $a$ and sons 
$T_{1}, ..., T_{n}$} is the tree $T$ of domain 
$D = \{<~>\}~\cup~\bigcup_{i} \{<i>\alpha~|~\alpha \in D_{i}\}$
such that
$$T(<~>) = a$$
and 
$$T(<i>\alpha) = T_{i}(\alpha)$$

If $T$ is a tree of domain $D$, $\alpha$ an occurrence of $D$ and $T'$ a tree 
of domain $D'$ then the {\it graft} of $T'$ in $T$ at the occurrence $\alpha$ 
($T[\alpha \la T']$) is the tree $T''$ of domain 
$D'' = D - \{\alpha \beta~|~\alpha \beta \in D\}
~\cup~\{\alpha \beta~|~\beta \in D'\}$ and such that 
$$T''(\gamma) = T'(\beta)~\mbox{if $\gamma = \alpha \beta$}$$
and 
$$T''(\gamma) = T(\gamma)~\mbox{otherwise}$$

Let $T$ and $T'$ be trees and let $a$ be a label such that all the
occurrences of
$a$ in $T$ are leaves $\alpha_{1}, ..., \alpha_{n}$ then the {\it substitution}
of $T'$ for $a$ in $T$ ($T[a \la T']$) is defined as 
$T[\alpha_{1} \la T'] ... [\alpha_{n} \la T']$. Note that since 
$\alpha_{1}, ..., \alpha_{n}$ are leaves, the order in which the 
grafts are performed is insignificant.
\end{Defs}

\subsection{Types}

\begin{Def} (Type)

Let us consider a finite set ${\cal T}$. The elements of ${\cal T}$ are called 
{\it atomic types}. A {\it type} is a tree whose labels are either the
elements of ${\cal T}$ or $\ra$ and such that the occurrences labeled with an 
element of ${\cal T}$ are leaves and the ones labeled with $\ra$ have two sons.

Let $T$ be a type, if the root of $T$ is labeled with an atomic type
$U$ then $T$ is written $U$, if the root of $T$ is labeled with $\ra$
and its sons are written $T_{1}$ and $T_{2}$
then $T$ is written $(T_{1} \ra T_{2})$. By convention
$T_{1} \ra T_{2} \ra T_{3}$ is an abbreviation for 
$(T_{1} \ra (T_{2} \ra T_{3}))$.
\end{Def}

\begin{Def} (Order of a Type)

If $T$ is a type, the {\it order} of $T$ is defined by\\
$\bullet$ $o(T) = 1$  if $T$ is atomic,\\
$\bullet$ $o(T_{1} \ra T_{2}) = max \{1+o(T_{1}),o(T_{2})\}$.
\end{Def}

\subsection{Typed $\lambda$-terms}

\begin{Defs}
For each type $T$ we consider three sets ${\cal C}_{T}$, 
${\cal I}_{T}$, ${\cal L}_{T}$. The elements
of ${\cal C}_{T}$ are called {\it constants} of type $T$, 
those of ${\cal I}_{T}$ {\it instantiable variables} of type $T$ and 
those of ${\cal L}_{T}$ {\it local variables} of type $T$.

We assume that we have in each atomic type at least a constant and
that there is a finite number of constants i.e. that the set
$\bigcup_{T} {\cal C}_{T}$ is finite
\footnote{This technical restriction is in fact superfluous, because
a matching problem expressed in a language with an infinite number of
constants can always be reduced to one expressed in the language
with a finite number of constants obtained by considering
only the constants occurring in the problem and one constant in each
atomic type.}.
We assume also that we have an infinite number of instantiable and
local variables of each type.

A typed $\lambda$-term is a tree whose labels are either $App$, or
$<Lam,x>$ where $x$ is a local variable, or $<Var,x>$ where $x$ is a 
constant, an instantiable variable or a local variable such that the 
occurrences labeled with $App$ have two sons, the occurrences labeled 
with $<Lam,x>$ have one son and the occurrences labeled with $<Var,x>$
are leaves.

Let $t$ be a term, if the root of $t$ is labeled with $<Var,x>$ we
write it $x$,
if the root of $t$ is labeled with $<Lam,x>$ and its son is written
$u$ then we write it 
$\lambda x:T . u$ where $T$ is the type of $x$, if the root of $t$ is
labeled with $App$ and its 
sons are written $u$ and $v$ then we write it $(u~v)$. By convention
$(u~v~w)$ is an abbreviation for $((u~v)~w)$.

In a term $t$, an occurrence $\alpha$ labeled with $<Var,x>$ is {\it bound} 
if there exists an occurrence $\beta$ in the path of $\alpha$ labeled with
$<Lam,x>$, it is {\it free} otherwise.

A term is {\it ground} if no occurrence is labeled with a pair $<Var,x>$ with 
$x$ instantiable.

Let $t$ and $t'$ be terms and $x$ be a variable, the {\it substitution} of $t'$
for $x$ in $t$ ($t[x \la t']$) is defined as $t[<Var,x> \la t']$. 
\end{Defs}

\begin{Def} (Type of a Term)

A term $t$ is said to have the type $T$ if either:\\
$\bullet$ $t$ is a constant, an instantiable variable or a local
variable of type $T$.\\
$\bullet$ $t = (u~v)$ and $u$ has type $U \ra T$ and $v$ type $U$ for some
type $U$,\\
$\bullet$ $t = \lambda x:U . u$, the term $u$ has type $V$ and $T = U \ra V$.

A term $t$ is said to be {\it well-typed} if there exists a type $T$ such that
$t$ has type $T$. In this case $T$ is unique and is called 
{\it the type of $t$}.
\end{Def}

\begin{Def} ($\beta \eta$-reduction)

The $\beta \eta$-reduction relation, written $\rhd$, is defined as the
smallest transitive relation compatible with term structure such that
$$(\lambda x:T . t~u) \rhd t[x \la u]$$
$$\lambda x:T.(t~x) \rhd t~~~\mbox{if $x$ is not free in $t$}$$
\end{Def}

We adopt the usual convention of considering terms up to $\alpha$-conversion
(i.e. bound variable renaming)
and we consider that bound variables are renamed to avoid capture during 
substitutions. A rigorous presentation would use, for instance, de
Bruijn indices \cite{deBruijn72}.

Obviously, if $t$ is a term of type $T$, $x$ is a variable of type $U$ and 
$u$ a term of type $U$ then the term $t[x \la u]$ has type $T$. In the same
way if a term $t$ has type $T$ and $t$ reduces to $u$ then $u$ has type $T$.

\begin{Prop}
The $\beta \eta$-reduction relation is strongly normalizable and 
confluent on typed terms, and thus each term has a unique normal form.

{\bd Proof} See, for instance, \cite{HinSel}.
\end{Prop}

\begin{Prop}
Let $t$ be a normal well-typed term of type
$U_{1} \ra ... \ra U_{n} \ra U~\mbox{($U$ atomic)}$,
the term $t$ has the form 
$$t = \lambda y_{1}:U_{1}.~...~\lambda y_{m}:U_{m}.(x~u_{1}~...~u_{p})$$
where $m \leq n$ and $x$ is a constant, an instantiable variable or a
local variable. 

{\bd Proof} The term $t$ can be written in a unique way
$t = \lambda y_{1}:V_{1}.~...~\lambda y_{m}:V_{m}.u$ where $u$ is not
an abstraction.
The term $u$ can be written in a unique way $u = (v~u_{1}~...~u_{p})$ where
$v$ is not an application. The term $v$ is not an application by definition,
it is not an abstraction (if $p = 0$ because $u$ is not an abstraction 
and if $p \neq 0$ because $t$ is normal), it is therefore a constant,
an instantiable variable or a local variable.
Then since $t$ has type $U_{1} \ra ... \ra U_{n} \ra U$, we have
$m \leq n$ and for all $i$, $V_{i} = U_{i}$.
\end{Prop}

\begin{Def} (Head of a Term, Atomic Term)

Let $t = \lambda y_{1}:T_{1}.~...~\lambda y_{m}:T_{m}.(x~u_{1}~...~u_{p})$
be a normal term. The symbol $x$ is called the {\it head} of the term.
If $m = 0$ then $t$ is said to be {\it atomic}, it is an abstraction otherwise.
\end{Def}

\begin{Def} ($\eta$-long Form)

If 
$t = \lambda y_{1}:U_{1}.~...~\lambda y_{m}:U_{m}.(x~u_{1}~...~u_{p})$
is a term of type 
$T = U_{1} \ra ... \ra U_{n} \ra U$ ($U$ atomic) ($m \leq n$)
which is in $\beta \eta$-normal form then we define its {\it $\beta$-normal 
$\eta$-long} form as the term
$$t' = \lambda y_{1}:U_{1}.~...~\lambda y_{m}:U_{m}. 
\lambda y_{m+1}:U_{m+1}.~...~\lambda y_{n}:U_{n}.
(x~u'_{1}~...~u'_{p}~y'_{m+1}~...~y'_{n})$$
where $u'_{i}$  is the
$\beta$-normal $\eta$-long form of $u_{i}$ and $y'_{i}$ is the $\beta$-normal
$\eta$-long form of $y_{i}$.

This definition is by induction on the pair $<c_{1},c_{2}>$ where $c_{1}$ is 
the number of occurrences in $t$ and $c_{2}$ the number of occurrences in $T$

In the following all the terms are assumed to be in $\beta$-normal $\eta$-long 
form.

\end{Def}

\subsection{B\"ohm Trees}

\begin{Def} (B\"ohm Tree)

A (finite) {\it B\"ohm tree} is a tree whose occurrences are labeled with
pairs $<l,x>$ such that $l$ is a list of local variables $<y_{1}, ... ,y_{n}>$ 
and $x$ is a constant, an instantiable variable or a local variable.
\end{Def}

\begin{Def} (Type of a B\"ohm Tree)

Let $t$ be a B\"ohm tree whose root is labeled with the pair 
$<<y_{1}, ... , y_{n}>,x>$
and whose sons are $u_{1}, ..., u_{p}$.
The B\"ohm tree $t$ is said to have the type $T$ if 
the B\"ohm trees $u_{1}, ..., u_{p}$ have type $U_{1}, ..., U_{p}$
the symbol $x$ has type $U_{1} \ra ... \ra U_{p} \ra U$ ($U$ atomic) and 
$T = T_{1} \ra ... \ra  T_{n} \ra U$ where $T_{1}, ..., T_{n}$ are the types
of the variables $y_{1}, ..., y_{n}$.

A B\"ohm tree $t$ is said to be {\it well-typed} if there exists a type $T$ 
such that $t$ has type $T$. In this case $T$ is unique and is called 
{\it the type of $t$}.
\end{Def}

\begin{Def} (B\"ohm Tree of a Normal Term)

Let $t = \lambda y_{1}:T_{1}.~...~\lambda y_{n}:T_{n}.(x~u_{1}~...~u_{p})$
be a $\lambda$-term in normal ($\eta$-long) form. 
The {\it B\"ohm tree} of $t$ is inductively defined as the tree whose root is 
the pair $<l,x>$ where $l = <y_{1}, ... ,y_{n}>$ is the list
of the variables bound at the top of this term, $x$ is the head symbol
of $t$ and sons are the B\"ohm trees of $u_{1}, ..., u_{p}$.
\end{Def}

\beginex{Remark}
Normal ($\eta$-long) well-typed terms and well-typed B\"ohm trees are
in one-to-one correspondence.
Moreover if $t$ is a normal ($\eta$-long) term and $\tilde{t}$ is its
B\"ohm tree
then occurrences in $t$ labeled with a constant, an instantiable
variable or a local variable and occurrences in
$\tilde{t}$ are in one-to-one correspondence.
So we will use the following abuse of notation: if $\alpha$ is an occurrence in
the B\"ohm tree of $t$ we write $(t/\alpha)$ for the normal
($\eta$-long) term 
corresponding to the B\"ohm tree $(\tilde{t}/\alpha)$ and $t[\alpha \la u]$ 
for the term $t[\alpha' \la u]$ where $\alpha'$ is the occurrence of a
variable or a constant in $t$ corresponding to $\alpha$.
\endex

\beginex{Notation}
Let $t$ be a term, we write $|t|$ for the depth of the B\"ohm tree of the
normal ($\eta$-long) form of $t$.
\endex

\begin{Prop}
In each type $T$ there is a ground term $t$ such that $|t| = 0$.

{\bd Proof} Let $T = U_{1} \ra ... \ra U_{n} \ra U$ with $U$ atomic and let 
$c$ be a constant of type $U$. The term 
$t = \lambda x_{1}:U_{1}.~...~\lambda x_{n}:U_{n}. c$ has type $T$ and
$|t| = 0$.
\end{Prop}

\subsection{Substitution}

\begin{Def} (Substitution)

A {\it substitution} is a finite set of pairs $<x_{i},t_{i}>$ where $x_{i}$
is an instantiable variable and $t_{i}$ a term of the same type in which no 
local variable occurs free such that if $<x,t>$ and $<x,t'>$ are both in this
set then $t = t'$.
The variables $x_{i}$ are said to be {\it bound} by the substitution.
\end{Def}

\begin{Def} (Substitution applied to a Term)

If $\sigma$ is a substitution and $t$ a term then we let
$$\sigma t = t[\alpha^{1}_{1} \la t_{1}] ... [\alpha^{p_{1}}_{1} \la t_{1}]
... [\alpha^{1}_{n} \la t_{n}] ... [\alpha^{p_{n}}_{n} \la t_{n}]$$
where $\alpha^{1}_{i}, ..., \alpha^{p_{i}}_{i}$ are the occurrences of 
$x_{i}$ in $t$.

Note that since the $\alpha^{j}_{i}$ are leaves, the order in which the 
grafts are performed is insignificant.
\end{Def}

\begin{Def} (Composition of Substitutions)

Let $\sigma$ and $\tau$ be two substitutions
the substitution $\tau \circ \sigma$ is defined by
$$\tau \circ \sigma = \{<x,\tau t>~|~<x,t> \in \sigma\} \cup 
\{<x,t>~|~<x,t> \in \tau~\mbox{and $x$ not bound by $\sigma$}\}$$
\end{Def}

\begin{Prop}
Let $\sigma$ and $\tau$ be two substitutions and $t$ is a term, 
we have 
$$(\tau \circ \sigma)t = \tau (\sigma t)$$

{\bd Proof} By decreasing induction on the depth of an occurrence $\alpha$ in
$t$ we prove that we have 
$$(\tau \circ \sigma)(t/\alpha) = \tau (\sigma (t/\alpha))$$
\end{Prop}

\section{Pattern Matching}

\begin{Def} (Matching Problem)

A {\it matching problem} is a set 
$\Phi = \{<a_{1}, b_{1}>, ..., <a_{n}, b_{n}>\}$ of pairs of terms
of the same type such that the terms $b_{1}, ..., b_{n}$ are ground. 
A pair $<a, b>$ is frequently written as an equation $a = b$.
\end{Def}

\begin{Def} (Third Order Matching Problem)

A {\it third order matching problem} is a matching problem 
$\Phi = \{a_{1} = b_{1}, ..., a_{n} = b_{n}\}$ such that the types
of the instantiable variables that occur in $a_{1}, ..., a_{n}$ are of
order at most three.
\end{Def}

\begin{Def} (Solution)

Let $\Phi = \{a_{1} = b_{1}, ..., a_{n} = b_{n}\}$ be a matching
problem. A substitution $\sigma$ is a {\it solution}
of this problem if and only if for every $i$, the normal form of the
terms $\sigma a_{i}$ and $b_{i}$ are identical up to $\alpha$-conversion.
\end{Def}

\beginex{Remark} Usual unification terminology distinguishes {\it variables}
(here instantiable variables) and {\it constants}.
The need for local variables comes from the fact that we want to transform the
problem $\lambda y:T.x = \lambda y:T.y$ (where $x$ is an instantiable
variable of type $T$)
into the problem $x = y$ by dropping the common abstraction. The symbol 
$y$ cannot be an instantiable variable (because it cannot be instantiated by
a substitution), it 
cannot be a constant because, if it were, we would have the solution
$x \la y$ to the second problem which is not a solution to the first.
So we let $y$ be a local variable and the solution $x \la y$ is now 
forbidden in both problems because no local variable can occur free in 
the terms substituted for variables in a substitution. 

In Huet's unification algorithm \cite{Huet75} \cite{Huet76} these local 
variables are always kept in the head of the terms in common abstractions.
In Miller's mixed prefixes terminology \cite{Miller91}, constants are 
universal variables declared to the left hand side of the instantiable
variables and local variables are universal variables declared to the
right hand side of all the instantiable variables.
\endex

\beginex{Remark} In an alternative definition of matching problems, the
terms $b_{1}, ..., b_{n}$ do not need to be ground.
The method of this paper can be adapted to such problems using the
standard technique of {\it variable freezing} \cite{Huet76}.
\endex

\begin{Def} (Ground Solution)

Let $\Phi = \{a_{1} = b_{1}, ..., a_{n} = b_{n}\}$ be a problem and
let $\sigma$ be a solution to $\Phi$.
The solution $\sigma$ is said to be {\it ground} if for each instantiable
variable that has an occurrence in some $a_{i}$, the term $\sigma x$ is ground.
\end{Def}

\begin{Prop}
If a matching problem has a solution then it has a ground solution.

{\bd Proof} 
Let $\Phi = \{a_{1} = b_{1}, ..., a_{n} = b_{n}\}$ be a matching
problem and let $\sigma$ be a solution to this problem. Let 
$y_{1}:T_{1}, ..., y_{n}:T_{n}$ be the instantiable variables
occurring in the 
term $\sigma x$ for some
$x$ instantiable variable occurring in some $a_{i}$. Let $u_{1}, ..., u_{n}$ be
ground terms of the types
$T_{1}, ..., T_{n}$. Let $\tau = \{<y_{1},u_{1}>, ..., <y_{n},u_{n}>\}$,
and $\sigma' = \tau \circ \sigma$. Obviously, for each instantiable variable 
$x$ of
$a$, the term $\sigma' x$ is ground and $\sigma'$ is a solution to $\Phi$.
\end{Prop}

\begin{Def} (Complete Set of Solutions)

Obviously if $\sigma$ is a solution to a problem $\Phi$ then for any 
substitution $\tau$, $\tau \circ \sigma$ is also a solution to $\Phi$.
A set $S$ of solutions to a problem $\Phi$
is said to be {\it complete} if for every substitution $\theta$ that
is a solution to
this problem there exists a substitution $\sigma \in S$ and a substitution
$\tau$ such that $\theta = \tau \circ \sigma$.
\end{Def}

\begin{Lem} Some problems have no finite complete set of solutions.

{\bd Proof (Example 1)}
Consider an atomic type $T$ and an instantiable variable
$x:T \ra (T \ra T) \ra T$ and the problem
$$\lambda a:T. (x~a~\lambda z:T.z) = \lambda a:T.a$$
The substitutions 
$$x \la \lambda o:T. \lambda s:T \ra T.(s~...~(s~o)~...~)$$
are solutions to this problem and they cannot be obtained as 
instances of a finite number of solutions.
\end{Lem}

\beginex{Remark} 
In \cite{Huet76} \cite{Zaionc}, the similar examples
$(x~\lambda z:T. z) = a$ and $(x~\lambda z:T. z) = b(a)$ are considered.
\endex

So in contrast with second order matching \cite{Huet76} \cite{HueLan} there is
no (always terminating) algorithm that enumerates a complete set of
solutions to a third order matching problem.

We consider now algorithms that take as an input a matching problem and either
give {\it one} solution to the problem or fail if it does not have any.

\section{A Bound on the Depth of Solutions}

All the problems considered in the rest of the paper are third order.

To prove the decidability of third order matching we are going to prove that 
the depth of the term $t$ substituted to a variable $x$ by a solution
$\sigma$ to a problem $\Phi$ can
be bounded by an integer $s$ depending only on the problem $\Phi$.
Of course the previous example shows that a matching problem may have 
solutions of arbitrary depth, but to design a decision algorithm we do not 
need to prove that {\it all} the solutions are bounded by $s$ but only that 
{\it at least one} is. 
To show this result we take a problem $\Phi$ that has a solution 
$\sigma$ (by proposition 5, we can consider without loss of generality that 
this solution is ground) and we build another solution $\sigma'$ whose depth is
bounded by an integer $s$ depending only on the problem $\Phi$.

The proof is divided into two parts. In the first part,we focus on a particular
case in which the problem $\Phi$ is a an {\it interpolation problem} i.e. 
set of equations of the form $(x~c_{1}~...~c_{n}) = b$ such
that $x$ is an instantiable variable and $c_{1}, ..., c_{n}$ and $b$
are ground terms. Then, in the second part, we reduce the general case to
this particular case.

Consider now an equation $(x~c_{1}~...~c_{n}) = b$ and a substitution
$\sigma$ solution to this equation.
Let us write
$t = \sigma x = \lambda y_{1}:T_{1}.~...~\lambda y_{n}:T_{n}.u$ ($u$ atomic).
We have 
$$\sigma (x~c_{1}~...~c_{n}) = 
(\lambda y_{1}:T_{1}.~...~ \lambda y_{n}:T_{n}.u~c_{1}~...~c_{n})$$
This term reduces to $u[y_{1} \la c_{1}, ..., y_{n} \la c_{n}]$ whose normal
form is $b$. 

The terms $c_{i}$ are at most second order.
In the key lemma, we prove that, {\it in the general case}, when we substitute
a second order term $c$ to a variable $y$ in a term $u$ and we
normalize the term $u[y \la c]$, we get a term with a depth larger
than or equal to the one of $u$. 
If this were true in all the cases, we would know that the depth
of $t$ (the solution) has to be less than or equal to the depth of $b$
(the right hand side of the equation). A simple enumeration of the terms
$t$ whose depth is less than or equal to $|b|$ would give a decision procedure.

Actually, the key lemma shows that the depth of the normal form of $u[y \la c]$
can be less than the depth of $u$ in two cases : when $c$ is a
non relevant term and when $|c| = 0$. When such cases happen, 
solutions may have an arbitrary depth. In these cases, we show that
if the problem $\Phi$ has a solution $\sigma$ then it has also another
solution $\sigma'$ whose depth is bounded by some integer $s$ depending 
only on the problem $\Phi$.

\subsection{Interpolation Problems}

\begin{Def} (Interpolation Problem)

An {\it interpolation problem} is a set of equations of the form 
$(x~c_{1}~...~c_{n}) = b$ such
that $x$ is an instantiable variable and $c_{1}, ..., c_{n}$ and $b$
are ground terms.
\end{Def}

\subsubsection{Key Lemma}

\begin{Def} (Relevant Term)

Let $c = \lambda z_{1}:U_{1}.~...~\lambda z_{p}:U_{p}. d$ 
($d$ atomic) be a normal term and $i$ an integer, $i \leq p$.
We say that $c$ is {\it relevant} in its $i^{th}$ argument if $z_{i}$ has an 
occurrence in the term $d$.
\end{Def}

\begin{Lem} (Key Lemma)
Let us consider a normal term $u$, a variable $y$ of type $T$ 
of order at most two and a normal ground term $c$ of type $T$.

(1) If $y$ has an occurrence in $u$ then $|c| \leq |u[y \la c]|$.

(2) If $\alpha$ is an occurrence in the B\"ohm tree of 
$u$ such that no occurrence in the path of $\alpha$ is labeled with $y$, 
then $\alpha$ is also an occurrence in the normal form of $u[y \la c]$ and 
has the same label in the B\"ohm tree of $u$ and in the B\"ohm tree of
the normal form of $u[y \la c]$.

(3) If $\alpha = <s_{1}, ..., s_{n}>$ is an occurrence in the B\"ohm tree 
of $u$ such that for each occurrence $\beta = <s_{1}, ..., s_{k}>$ in the path
of $\alpha$, $\beta \neq \alpha$, labeled with $y$, the term $c$ is relevant in 
its 
$r^{th}$ argument where $r$ is the position of the son of $\beta$ in the path 
of $\alpha$ i.e. $r = s_{k + 1}$, then there exists an occurrence $\alpha'$ 
of the B\"ohm tree of the normal form of $u[y \la c]$ such that all the labels
occurring in the path of $\alpha$,
except $y$, occur in the path of $\alpha'$ and the number of times they
occur in the path of $\alpha'$ is greater than or equal to the number of times 
they occur in the path of $\alpha$. Moreover if the occurrence 
$\alpha$ is labeled with a symbol different from $y$, then the occurrence
$\alpha'$ is labeled with this same symbol.

(4) Moreover if $|c| \neq 0$ then the length of $\alpha'$ is greater
than or equal to the length of $\alpha$.

{\bd Proof} By induction on the number of occurrences of $y$ in $u$.
We substitute these occurrences one by one and we normalize the term. Let 
$\beta$ be the occurrence in the B\"ohm tree of $u$ corresponding to
the occurrence of $y$ in $u$ we substitute.
Let us write 
$$c = \lambda z_{1}:U_{1}.~...~\lambda z_{p}:U_{p}. d$$
The term $(u/\beta)$ has the form 
$\lambda v_{1}:V_{1}.~...~\lambda v_{q}:V_{q}.(y~e_{1}~...~e_{p})$. 
When we substitute $y$ by the term $c$ in $(y~e_{1}~...~e_{p})$ we get 
$(c~e_{1}~...~e_{p})$ and when we normalize this term we get the term
$d[z_{1} \la e_{1}, ..., z_{p} \la e_{p}]$ which is normal
because the type of the $e_{i}$ are first order.

Let us consider the occurrences in the B\"ohm tree of
$u$, while substituting the occurrence of $y$ corresponding to $\beta$,
we have removed all the occurrences $\beta <i> \gamma$ where
$i$ is an integer ($i \leq p$) and $\gamma$ is an occurrence in the B\"ohm 
tree of $e_{i}$.
We have added all the occurrences $\beta \delta$ where $\delta$ is an 
occurrence of the B\"ohm tree of $c$ labeled with a symbol different from 
$z_{1}, ..., z_{p}$ and all the occurrences $\beta \delta \gamma$ where 
$\delta$ is a leaf occurrence in the B\"ohm tree of $c$ 
labeled with a $z_{i}$ and $\gamma$ is an occurrence of the B\"ohm tree of 
$e_{i}$.

\begin{center}
\begin{picture}(230,143)(0,0)

\put(0,0) {\framebox(230,143)}

\put (40,133) {\line (-1,-2) {20}}
\put (40,133) {\line (1,-2) {20}}
\put (20,93) {\line (1,0) {40}}

\put (42,86) {\footnotesize $\beta$}

\put (40,93) {\line (-1,-5) {2}}

\put (42,78) {\footnotesize $\beta<i>$}

\put (38,83) {\line (-1,-2) {20}}
\put (38,83) {\line (1,-2) {20}}
\put (18,43) {\line (1,0) {40}}
\put (33,53) {$e_{i}$}

\put (21,34) {\footnotesize $\beta <i> \gamma$}

\put (80,70) {\vector (1,0){70}}

\put (190,133) {\line (-1,-2) {20}}
\put (190,133) {\line (1,-2) {20}}
\put (170,93) {\line (1,0) {40}}

\put (195,86) {\footnotesize $\beta$}

\put (190,93) {\line (-1,-2) {20}}
\put (190,93) {\line (1,-2) {20}}
\put (170,53) {\line (1,0) {40}}
\put (187,63) {$c$}

\put (189,46) {\footnotesize $\beta \delta$}

\put (185,53) {\line (-1,-2) {20}}
\put (185,53) {\line (1,-2) {20}}
\put (165,13) {\line (1,0) {40}}
\put (180,23) {$e_{i}$}

\put (178,4) {\footnotesize $\beta \delta \gamma$}

\end{picture}
\end{center}

(1) Let $\beta$ be an outermost occurrence of $y$ in the B\"ohm tree of $u$. 
For each 
occurrence $\delta$ in the B\"ohm tree of $c$, $\beta\delta$ is an occurrence 
in the B\"ohm tree of the normal form of $u[y \la c]$. So 
$|c| \leq |u[y \la c]|$.

(2) When an occurrence $\beta$ of $y$ is substituted by $c$ all the occurrences
removed have the form $\beta <i> \gamma$. So if no occurrence in the path of
$\alpha$ is labeled with $y$, the occurrence $\alpha$ remains in the
normal form of $u[y \la c]$.

(3) If the occurrence $\beta$ is not in the path of $\alpha$ then the 
occurrence $\alpha$ is still an occurrence in the normal form of 
$u[y \la c]$, we take $\alpha' = \alpha$.

If $\beta = \alpha$ then the occurrence $\beta$ is an occurrence of the
B\"ohm tree of the normal form of $u[y \la c]$. We take 
$\alpha' = \beta = \alpha$.

If $\beta$ is in the path of $\alpha$ and $\beta \neq \alpha$, 
$\beta = <s_{1}, ..., s_{k}>$ then let $r$ be the position of the son of 
$\beta$ in the path of $\alpha$ i.e. $r = s_{k + 1}$.
Let $\gamma$ be such that $\alpha = \beta<r>\gamma$.
By hypothesis $z_{r}$ has an occurrence in $d$, let $\delta$ be such an  
occurrence.
The occurrence $\beta \delta \gamma$ is an occurrence in the B\"ohm tree of 
the normal form of $u[y \la c]$. We take $\alpha' = \beta \delta \gamma$.

In all the cases, all the labels occurring in the path of $\alpha$, except $y$,
occur in the path of $\alpha'$ and the number of times they occur in the
path of $\alpha'$ is greater than or equal to the number of times they
occur in the path of $\alpha$.

If the occurrence $\alpha$ is labeled with a symbol different from $y$, then 
the occurrence $\alpha'$ is labeled with the same symbol as $\alpha$.

(4) If $\delta = <~>$ then 
$c = \lambda z_{1}:U_{1}.~...~\lambda z_{p}:U_{p}. z_{r}$ and 
$|c| = 0$. So if $|c| \neq 0$ then $\delta \neq <~>$ and the length of 
$\alpha'$ is greater than or equal to the length of $\alpha$.
\end{Lem}

\beginex{Corollary}
Let us consider a normal term $u$, a variable $y$ of type $T$ of 
order at most two and a ground term $c$ of type $T$.
If $c$ is relevant in all its arguments and $|c| \neq 0$ then 
$|u| \leq |u[y \la c]|$.

{\bd Proof}
We take for $\alpha$ the longest occurrence in the B\"ohm tree of $u$.
When we substitute one by one the occurrences of $y$, by part (4) of the key 
lemma, we get occurrences that are at least long.
So there is an occurrence in the B\"ohm tree of the normal form of $u[y \la c]$
which is at least long as $\alpha$. So $|u| \leq |u[y \la c]|$.
\endex

\subsubsection{Computing the Substitution $\sigma'$}

Let us consider an equation $(x~c_{1}~...~c_{n}) = b$. Let
$\sigma$ be a solution to this equation and let $t = \sigma x$. Let us write
$t = \lambda y_{1}:T_{1}.~...~\lambda y_{n}:T_{n}. u$.
The normal form of the term $\sigma (x~c_{1}~...~c_{n})$ is the normal form of 
$u[y_{1} \la c_{1}, ..., y_{n} \la c_{n}]$. If all the 
$c_{i}$ are relevant in their arguments and $|c_{i}| \neq 0$ 
then using the corollary of the key lemma we have
$|t| \leq |(t~c_{1}~...~c_{n})|$, so $|t| \leq |b|$ and this gives a 
bound on the depth of $t$.
But the depth of $t$ may decrease when applied to the terms $c_{i}$ and 
normalized in two cases:\\
$\bullet$ if one of the terms $c_{i}$ is not relevant in one of its 
arguments,\\
$\bullet$ if one of the terms $c_{i}$ is such that $|c_{i}| = 0$.\\
So solutions may have an arbitrary depth.
When this happens, we compute another solution to the problem 
whose depth is bounded by an integer $s$ depending only on the initial problem.

This new substitution is constructed in two steps. In the first step
we deal with non relevant terms and in the second with terms of depth 0.

\beginex{Example 2}
Let $x$ be an instantiable variable of type $T \ra (T \ra T) \ra T$.
Consider the problem 
$$(x~a~\lambda z:T.b) = b$$
The variable $z$ has no occurrence in $b$ so this problem has solutions of 
arbitrary depth 
$$x \la \lambda o:T. \lambda s:T \ra T. (s~t)$$
where $t$ is an arbitrary term of type $T$. In this example we will compute 
the substitution 
$$x \la  \lambda o:T. \lambda s:T \ra T.(s~c)$$
where $c$ is a constant.
\endex

\beginex{Example 1 (continued)} The term $\lambda z:T.z$ has depth 0, so we
have solutions of an arbitrary depth. In this example we will compute the 
substitution 
$$x \la \lambda o:T. \lambda s:T \ra T.(s~o)$$

\endex

\begin{Def} (Occurrence Accessible with Respect to an Equation of the
Form $(x~c_{1}~...~c_{n}) = b$)

Let us consider an equation 
$$(x~c_{1}~...~c_{n}) = b$$
and the term
$$t = \sigma x = \lambda y_{1}:T_{1}.~...~\lambda y_{n}:T_{n}. u$$
Let us consider the B\"ohm tree of $t$. The set of the occurrences of the 
B\"ohm tree of $t$ {\it accessible} with respect to
the equation $(x~c_{1}~...~c_{n}) = b$ is inductively defined as:\\
$\bullet$ the root of the B\"ohm tree of $t$ is accessible,\\
$\bullet$ if $\alpha$ is an accessible occurrence labeled with $y_{i}$ and 
$c_{i}$ is relevant in its $j^{th}$ argument then the occurrence $\alpha<j>$ 
(the $j^{th}$ son of $\alpha$) is accessible,\\
$\bullet$ if $\alpha$ is an accessible occurrence labeled with a
symbol different from all the $y_{i}$ then all the sons of $\alpha$
are accessible.
\end{Def}

\begin{Def} (Occurrence Accessible with Respect to an Interpolation
Problem)

An occurrence is {\it accessible} with respect to an interpolation problem
if it is accessible with respect to one of the equations of this problem.
\end{Def}

\begin{Def} (Term Accessible with Respect to an Interpolation Problem)

A term is {\it accessible} with respect to an interpolation problem
if all the occurrences of its B\"ohm tree which are not leaves are
accessible with respect to this problem.
\end{Def}

\begin{Def} (Accessible Solution Built from a Solution)

Let $\Phi$ be an interpolation problem and let $\sigma$ be a solution
to this problem.
For each instantiable variable $x$ occurring in the equations of $\Phi$
we consider the term $t = \sigma x$.
In the B\"ohm tree of $t$, we prune all the occurrences non accessible 
with respect to the equations of $\Phi$ in which $x$ has an 
occurrence and put B\"ohm trees of ground terms of depth 0 of the expected 
type as leaves. The tree obtained that way
is the B\"ohm tree of some term $t'$. We let $\hat{\sigma} x = t'$.
\end{Def}

\beginex{Example 2 (continued)}
From the solution 
$$x \la \lambda o:T. \lambda s:T \ra T.(s~t)$$ 
where $t$ is an 
arbitrary term, we compute the substitution 
$$x \la \lambda o:T. \lambda s:T \ra T. (s~c)$$
where $c$ is a constant.
\endex

\begin{Prop}
Let $\Phi$ be an interpolation problem and let $\sigma$ be a solution
to $\Phi$, then the accessible solution $\hat{\sigma}$ built from
$\sigma$ is a solution to $\Phi$.

{\bd Proof} 
Let us consider an equation $(x~c_{1}~...~c_{n}) = b$ of $\Phi$ and the terms 
$$\sigma x = t = \lambda y_{1}:T_{1}.~...~\lambda y_{n}:T_{n}. u$$
and 
$$\hat{\sigma} x = t' = \lambda y_{1}:T_{1}.~...~\lambda y_{n}:T_{n}. u'$$
We prove by decreasing induction on the depth of the occurrence $\alpha$ of
the B\"ohm tree of $u$ that if $\alpha$ is accessible with respect to the
equation $(x~c_{1}~...~c_{n}) = b$
then $\alpha$ is also an occurrence of the B\"ohm tree of $u'$ and
$$(u'/\alpha)[y_{1} \la c_{1}, ..., y_{n} \la c_{n}] = 
(u/\alpha)[y_{1} \la c_{1}, ..., y_{n} \la c_{n}]$$
and then since the root of $u$ is accessible with respect to this equation
we have
$$u'[y_{1} \la c_{1}, ..., y_{n} \la c_{n}] =
u[y_{1} \la c_{1}, ..., y_{n} \la c_{n}]$$
i.e.
$$((\hat{\sigma}x)~c_{1}~...~c_{n}) = b$$
So $\hat{\sigma}$ is a solution to $\Phi$.
\end{Prop}

\begin{Prop} 
Let $\Phi$ be an interpolation problem and let $\sigma$ be a solution
to $\Phi$. Let $h$ be the maximum
depth of the right hand side of the equations of $\Phi$.
Let $\hat{\sigma}$ the accessible solution built from $\sigma$. Let
$$t = \hat{\sigma} x = \lambda y_{1}:T_{1}.~...~\lambda y_{n}:T_{n}.u$$ 
($u$ atomic). There are at most $h+1$ occurrences of symbols not in 
$\{y_{1}, ..., y_{n}\}$ on a path of the B\"ohm tree of $t$.

{\bd Proof} 
Let $\alpha$ be an occurrence in the B\"ohm tree of $t$ 
such that there are more than $h+1$ occurrences of symbols not in 
$\{y_{1}, ..., y_{n}\}$ in the path of $\alpha$. 

Let $\beta$ be the $(h+1)-th$ occurrence of such a symbol. Since there are more
that $h+1$ occurrences of symbols not in $\{y_{1}, ..., y_{n}\}$ in the path
of $\alpha$, the occurrence $\beta$ is not a leaf, so it is accessible with 
respect to some equation $(x~c_{1}~...~c_{n}) = b$ of $\Phi$. Also, since this 
occurrence is not a leaf, it is labeled with a symbol $f$ whose type
is not first order. 

For each occurrence $\gamma = <s_{1}, ..., s_{k}>$ in the path of $\beta$ 
labeled with $y_{i}$, let $r$ be the position of the son of this occurrence in 
this path (i.e. $r = s_{k+1}$). Since the occurrence $\beta$ is accessible 
with respect to the equation $(x~c_{1}~...~c_{n}) = b$, the term $c_{i}$ is 
relevant in its $r^{th}$ argument.
So using $n$ times the part (3) of the key lemma there exists an
occurrence $\beta'$ in the B\"ohm tree of the normal form of the term
$b = (\hat{\sigma} x~c_{1}~...~c_{n})$ such that the path of $\beta'$ contains
at least $h+1$ occurrences. Thus, the length of this occurrence is at
least $h$. 
This occurrence is labeled with the symbol $f$ whose type is not first
order, so it has a son $\beta''$ whose length is at least $h+1$. 

So the depth of $b$ is greater than or equal to $h+1$ which is contradictory.
\end{Prop}

\begin{Def} (Compact Term)

A term $t = \lambda y_{1}:T_{1}.~...~\lambda y_{n}:T_{n}.u$ ($u$ atomic) 
is {\it compact}
with respect to an interpolation problem $\Phi$ if no variable $y_{i}$ has
more than $h+1$ occurrences in a path of its B\"ohm tree, where $h$ is the
maximum depth of the right hand side of the equations of $\Phi$.
\end{Def}

\begin{Prop}
Let $\Phi$ be an interpolation problem 
and let $\hat{\sigma}$ be an accessible solution to $\Phi$. Let
$h$ be the maximum depth of the right hand side of the equations of $\Phi$.
Let us consider an instantiable variable $x$ and 
$$t = \hat{\sigma} x = \lambda y_{1}:T_{1}.~...~\lambda y_{n}:T_{n}. u$$
($u$ atomic). Let us consider a variable $y_{i}$ and an
occurrence $\alpha$ of the B\"ohm tree of $t$ such that there are more than
$h+1$ occurrences on the path of $\alpha$ labeled with the variable $y_{i}$. 

We consider all the equations $(x~c_{1}~...~c_{n}) = b$ of $\Phi$ such that the
$(h+2)-th$ occurrence of $y_{i}$ is accessible with respect to this equation.
Then there exists an integer $j$ such that for every such equation 
we have 
$$c_{i} = \lambda z_{1}:U_{1}.~...~\lambda z_{p}:U_{p}. z_{j}$$

{\bd Proof}
Let $\beta$ be the first occurrence of $y_{i}$ in the path of $\alpha$. Let 
$j$ be the integer such that\\
$\alpha = \beta <j> \beta'$. 

Let $(x~c_{1}~...~c_{n}) = b$ be an equation of $\Phi$ such that the $(h+2)-th$
occurrence of $y_{i}$ on the considered path is accessible with respect to 
this equation.

If the head of $c_{i}$ is a symbol different from a $z_{k}$ then
$|c_{i}| \neq 0$. Using part (3) of the key lemma
when we substitute $c_{1}, ..., c_{i-1}$, $c_{i+1}$, ..., $c_{n}$ we have 
an occurrence $\alpha'$ that has more than $h+1$ occurrences of $y_{i}$ on 
its path. Then using part (4) of the key lemma, when we substitute $c_{i}$
we have an occurrence $\alpha''$ whose length is greater than or equal
to $h+1$ so 
$$h+1 \leq |u[y_{1} \la c_{1}, ..., y_{n} \la c_{n}]|$$ 
i.e. $h+1 \leq |b|$ which is contradictory. So we have 
$$c_{i} = \lambda z_{1}:U_{1}.~...~\lambda z_{p}:U_{p}. z_{k}$$

Since $h+2 > 1$ the occurrence $\beta<j>$ is accessible with respect to
the equation $(x~c_{1}~...~c_{n}) = b$.
Thus as the occurrence $\beta$ is labeled with $y_{i}$
and the occurrence 
$\beta<j>$ is accessible with respect to this equation, the term $c_{i}$ is 
relevant in its $j^{th}$ argument. Therefore $k = j$ and
$$c_{i} = \lambda z_{1}:U_{1}.~...~\lambda z_{p}:U_{p}. z_{j}$$
\end{Prop}

\begin{Def} (Compact Accessible Solution Built from an Accessible
Solution)

Let $\Phi$ be an interpolation problem and let $\hat{\sigma}$ be an 
accessible solution to this problem. 
Let $h$ be the maximum depth of a right hand side of the equations of $\Phi$. 
We let 
$$\hat{\sigma} x = t = \lambda y_{1}:T_{1}.~...~\lambda y_{n}:T_{n}. u$$
For each $\alpha$, occurrence in $t$ labeled with $y_{i}$ such that the
corresponding occurrence $\alpha'$ in the B\"ohm tree of $t$ has more than
$h+1$ occurrences labeled with $y_{i}$ in its path, 
we have $c_{i} = \lambda z_{1}:U_{1}.~...~\lambda z_{p}:U_{p}. z_{j}$ in 
all the equations $(x~c_{1}~...~c_{n}) = b$ of $\Phi$ such 
that $\alpha'$ is accessible with respect to this equation.
We substitute the occurrence $\alpha$ by the term 
$\lambda z_{1}:U_{1}.~...~\lambda z_{p}:U_{p}. z_{j}$.
We get that way a term $t'$. We let $\sigma' x = t'$.
\end{Def}

\beginex{Example 1 (continued)}
We build the substitution 
$$x \la \lambda o:T. \lambda s:T \ra T. (s~o)$$
\endex

\beginex{Example 3} Consider an instantiable variable $x$ of type 
$(T \ra T \ra T) \ra T$. And the problem
$$(x~\lambda y:T. \lambda z:T. y) = a$$
$$(x~\lambda y:T. \lambda z:T. z) = b$$
We have the solution 
$$x \la \lambda f:T \ra T \ra T.(f~a~(f~c~(f~d~b)))$$

This solution is accessible but not compact. The first occurrence of $f$ is 
accessible with respect to both equations, but the second and third 
occurrences are accessible only with respect to the second one. We have 
$h = 0$, so we substitute the second and third occurrences of $f$ by the term 
$\lambda y:T. \lambda z:T. z$ and we get the substitution 
$$x \la \lambda f:T \ra T \ra T. (f~a~b)$$
Note that we must not substitute the first occurrence of $f$ by
$\lambda y:T. \lambda z:T. z$,
because we would get the substitution  $x \la \lambda f:T \ra T \ra T. b$ 
which is not a solution to the first equation.
\endex

\begin{Prop} Let $\Phi$ be an interpolation problem and let $\sigma$
be a solution to $\Phi$.
Let $\hat{\sigma}$ the accessible solution built from $\sigma$ and 
$\sigma'$ the compact accessible solution built from $\hat{\sigma}$.
Then $\sigma'$ is a solution to $\Phi$.

{\bd Proof} 
We consider an equation $(x~c_{1}~...~c_{n}) = b$ and we let 
$$\hat{\sigma} x = t = \lambda y_{1}:T_{1}.~...~\lambda y_{n}:T_{n}. u$$
and
$$\sigma' x = t = \lambda y_{1}:T_{1}.~...~\lambda y_{n}:T_{n}. u'$$

The term $u'$ is obtained by substituting in the term $u$ some occurrences 
(say $\beta_{1}, ..., \beta_{k}$) by some terms (say $e_{1}, ..., e_{k}$).
If $\alpha$ is an occurrence of $u$ then we define $u'_{\alpha}$ as the term
obtained by substituting in the term $u/\alpha$ the occurrence $\gamma_{i}$
by the term $e_{i}$ if $\beta_{i} = \alpha \gamma_{i}$.

We prove by decreasing induction on the depth of the occurrence $\alpha$ of
the B\"ohm tree of $u$ that if $\alpha$ is accessible with respect to the
equation $(x~c_{1}~...~c_{n}) = b$
then
$$(u'_{\alpha})[y_{1} \la c_{1}, ..., y_{n} \la c_{n}] = 
(u/\alpha)[y_{1} \la c_{1}, ...,  y_{n} \la c_{n}]$$
Thus for the root we get
$$u'[y_{1} \la c_{1}, ..., y_{n} \la c_{n}] =
u[y_{1} \la c_{1}, ..., y_{n} \la c_{n}]$$
i.e.
$$((\sigma' x)~c_{1}~...~c_{n}) = b$$
So $\sigma'$ is a solution to all the equations of $\Phi$.
\end{Prop}

\begin{Prop} Let $\Phi$ be an interpolation problem and let $\sigma$
be a solution to $\Phi$. Let $\hat{\sigma}$ be the accessible solution 
built from $\sigma$ and $\sigma'$ the compact accessible solution built from 
$\hat{\sigma}$.
Let $h$ be the maximum depth of the right hand side of the equations of $\Phi$.
For every instantiable variable $x$ of arity $n$, $\sigma' x$ has a depth less
than or equal to $(n+1)(h+1)-1$.

{\bd Proof} In a path of the B\"{o}hm tree of $\sigma' x$ each $y_{i}$ has at 
most $h+1$ occurrences and there are at most $h+1$ occurrences of other 
symbols, so there are at most $(n+1)(h+1)$ occurrences. Therefore the depth 
of $\sigma' x$ is bounded by $(n+1)(h+1)-1$.
\end{Prop}

\begin{Lem} Let $\Phi$ be a third order interpolation problem.
If $\Phi$ has a solution $\sigma$ then it also has a
solution $\sigma'$ such that for every instantiable variable $x$,
$\sigma x$ has a depth 
less than or equal to $(n+1)(h+1)-1$, where $h$ is maximum of the
depths of the right hand side of the equations and $n$ the arity of $x$.

{\bd Proof} The compact accessible solution $\sigma'$ built from the 
accessible solution built from the solution $\sigma$ is a solution and 
for every instantiable variable $x$, $\sigma' x$ has a depth less
than or equal to $(n+1)(h+1)-1$.

This bound is met, for instance by the example 3. 
\end{Lem}

\subsection{General Case}

Let $a = b$ be an equation and let $\sigma$ be a 
solution to this equation. We construct an interpolation problem 
$\Phi(a=b,\sigma)$ such that for every equation 
$(x~c_{1}~...~c_{n}) = b'$ of $\Phi(a=b,\sigma)$ we have 
$|b'| \leq |b|$, $\sigma$ is a solution to  $\Phi(a=b,\sigma)$ and
every solution to $\Phi(a=b,\sigma)$ is a solution to $a = b$.

\begin{Def} Let $a = b$ be an equation and let $\sigma$ be a (ground)
solution to this equation. By induction on the number of occurrences of
$a$ we construct an interpolation problem $\Phi(a = b,\sigma)$.

$\bullet$ If $a = \lambda x:T. d$ then since $\sigma$ is a solution to
the problem 
$a = b$ we have $b = \lambda x:T.e$ and $\sigma$ is a solution to the problem 
$d = e$. We let 
$$\Phi(a = b,\sigma) = \Phi(d = e,\sigma)$$

$\bullet$ If $a = (f~d_{1}~...~d_{n})$ with $f$ a constant or a local
variable then since 
$\sigma$ is a solution to $a = b$ we have $b = (f~e_{1}~...~e_{n})$
and $\sigma$ is a solution to the problems $d_{i} = e_{i}$. We let
$$\Phi(a = b,\sigma) = \bigcup_{i} \Phi(d_{i} = e_{i},\sigma)$$

$\bullet$ If $a = (x~d_{1}~...~d_{n})$ with $x$ instantiable then for
all $i$ such that 
$z$ has an occurrence in the normal form of the term
$(\sigma x~\sigma d_{1}~...~\sigma d_{i-1}~z~\sigma d_{i+1}~...~\sigma d_{n})$
we let $c_{i} = \sigma d_{i}$ and 
$H_{i} = \Phi(d_{i} = \sigma d_{i},\sigma)$ (obviously $\sigma$ is
a solution to $d_{i} = \sigma d_{i}$).
Otherwise we let $c_{i} = z_{i}$ where $z_{i}$ is a new local variable and
$H_{i} = \emptyset$. We let 
$$\Phi(a = b,\sigma) = \{(x~c_{1}~...~c_{n}) = b\} \cup \bigcup_{i} H_{i}$$
\end{Def}

\begin{Prop}
Let $t = (x~d_{1}~...~d_{n})$ be a term and let $\sigma$ be a substitution.
Let $c_{i} = \sigma d_{i}$ if $z$ has an occurrence in 
$(\sigma x~\sigma d_{1}~...~\sigma d_{i-1}~z~\sigma d_{i+1}~...~\sigma d_{n})$
and $c_{i} = z_{i}$ where $z_{i}$ is a new local variable of the same
type as $d_{i}$ otherwise.
The variables $z_{i}$ do not occur in the normal form of
$(\sigma x~c_{1}~...~c_{n})$.

{\bd Proof} Let us assume that some of these variables have an occurrence 
in the normal form of $(\sigma x~c_{1}~...~c_{n})$
and consider an outermost 
occurrence of such a variable $z_{i}$ in the B\"ohm tree
of the normal form of $(\sigma x~c_{1}~...~c_{n})$.
By part (2) of the key lemma, the variable $z_{i}$ has also an occurrence 
in the normal form of term
$(\sigma x~c_{1}~...~c_{n})[z_{j} \la \sigma d_{j}~|~j \neq i]$ i.e.
in the normal form of the term
$(\sigma x
~\sigma d_{1}~...~\sigma d_{i-1}~z_{i}~\sigma d_{i+1}~...~\sigma d_{n})$,
which is contradictory.
\end{Prop}

\begin{Prop} 
Let $a = b$ be an equation and let $\sigma$ be a solution to this equation,

$\bullet$ the substitution $\sigma$ is a solution to $\Phi(a = b,\sigma)$,

$\bullet$ conversely, if $\sigma'$ is a solution to 
$\Phi(a = b,\sigma)$ then $\sigma'$ is also a solution to the equation $a = b$.

{\bd Proof} 

$\bullet$ By induction on the number of occurrences of $a$. When $a$ is an 
abstraction $a = \lambda x:T.d$ (resp. an atomic term whose head is a 
constant or local variable $a = (f~d_{1}~...~d_{n})$)
then $b$ is also an abstraction $b = \lambda x:T.e$ (reps. an atomic
term with the same head $b = (f~e_{1}~...~e_{n})$)
and by induction hypothesis $\sigma$ is a solution to all the equations of
the set $\Phi(d=e,\sigma)$ (resp. $\Phi(d_{i} = e_{i},\sigma)$), so it is a
solution to all the equations of $\Phi (a = b,\sigma)$.

When $a = (x~d_{1}~...~d_{n})$ then by induction hypothesis $\sigma$ is a 
solution to all the equations of the sets $H_{i}$ and using the previous 
proposition the variables $z_{i}$ have no occurrences in the term 
$(\sigma x~c_{1}~...~c_{n})$ so we have
$$(\sigma x~c_{1}~...~c_{n}) =
(\sigma x~c_{1}~...~c_{n})[z_{i} \la \sigma d_{i}]$$
$$(\sigma x~c_{1}~...~c_{n}) = (\sigma x~\sigma d_{1}~...~\sigma d_{n}) = b$$
So $\sigma$ is a solution to the equation $(x~c_{1}~...~c_{n}) = b$.

$\bullet$ By induction on the number of occurrences of $a$. 
Let $\sigma'$ be a substitution solution to 
$\Phi(a = b,\sigma)$. If $a$ is an abstraction $a = \lambda x:T.d$ 
(resp. an atomic term whose head is a constant or a local variable
$a = (f~d_{1}~...~d_{n})$) 
then $b$ is also an abstraction $b = \lambda x:T.e$ (reps. an atomic
term with the same head $b = (f~e_{1}~...~e_{n})$)
and by induction hypothesis we have 
$\sigma' d = e$ (resp. $\sigma' d_{i} = e_{i}$) and so $\sigma' a = b$.

If $a = (x~d_{1}~...~d_{n})$ then we have 
$$(\sigma' x~c_{1}~...~c_{n}) = b$$
and for all $i$ such that $z$ has an occurrence in 
$(\sigma x~\sigma d_{1}~...~\sigma d_{i-1}~z~\sigma d_{i+1}~...~\sigma d_{n})$
by induction hypothesis we have $\sigma' d_{i} = \sigma d_{i}$,
so $c_{i} = \sigma' d_{i}$. Therefore
$$(\sigma' x~c_{1}~...~c_{n})[z_{i} \la \sigma' d_{i}] 
= b[z_{i} \la \sigma' d_{i}]$$
$$(\sigma' x~c_{1}~...~c_{n})[z_{i} \la \sigma' d_{i}] = b$$
$$(\sigma' x~\sigma' d_{1}~...~\sigma' d_{n}) = b$$
$$\sigma' a = b$$
\end{Prop}

\begin{Prop}
Let $a = b$ be an equation and let $\sigma$ be a solution to this equation,
if $a' = b'$ is an equation of $\Phi(a = b,\sigma)$ then $|b'| \leq |b|$.

{\bd Proof} 
By induction on the number of occurrences of $a$. When 
$a$ is an abstraction  $a = \lambda x:T.d$ (reps. an atomic term whose
head is a constant or a local variable $a = (f~d_{1}~...~d_{n})$)
then $b$ is also an abstraction $b = \lambda x:T.e$ (reps. an atomic
term with the same head $b = (f~e_{1}~...~e_{n})$)
and by induction hypothesis $|b'| \leq |e|$ (resp. $|b'| \leq |e_{i}|$) so
$|b'| \leq |b|$.

When $a = (x~d_{1}~...~d_{n})$ and the considered equation is
$(x~c_{1}~...~c_{n}) = b$ then we have $b' = b$ so $|b'| \leq |b|$.
When the considered equation is in one of the sets $H_{i}$,
the set $H_{i}$ is non empty so $z$ has an occurrence in the normal form of
the term 
$(\sigma x~\sigma d_{1}~...~\sigma d_{i-1}~z~\sigma d_{i+1}~...~\sigma d_{n})$
and 
$(\sigma x~\sigma d_{1}~...~\sigma d_{i-1}~z~\sigma d_{i+1}~...~\sigma d_{n})
[z \la \sigma d_{i}] = b$
so using part (1) of the key lemma we have $|\sigma d_{i}| \leq |b|$
and by induction hypothesis $|b'| \leq |\sigma d_{i}|$ so $|b'| \leq |b|$.
\end{Prop}

\begin{Def}
Let $\Psi$ be a third order matching problem and let $\sigma$ be a
solution to $\Psi$. We let 
$\Phi(\Psi,\sigma)$ be the following third order interpolation problem:
$$\Phi(\Psi,\sigma) = \bigcup_{a=b \in \Psi} \Phi(a=b,\sigma)$$
\end{Def}

\begin{Prop}
Let $\Psi$ be a third order matching problem and let $\sigma$ be a 
solution to $\Psi$. Let $h$ be
the maximum of the depth of the right hand side of the equations of
$\Psi$. Then $\sigma$ is a
solution to the problem $\Phi(\Psi,\sigma)$, each substitution
$\sigma'$ solution to the problem $\Phi(\Psi,\sigma)$ is a solution to
$\Psi$ and if $a' = b' \in \Phi(\Psi,\sigma)$ then $|b'| \leq h$.

{\bd Proof} By propositions 12 and 13.
\end{Prop}

\begin{Lem} Let $\Psi$ be third order matching problem. 
Let $h$ be the maximum of the depth of the the right hand side of the
equations of $\Psi$.
If this problem has a solution $\sigma$ then it also has a
solution $\sigma'$ such that for every instantiable variable $x$,
$\sigma x$ has a depth less than or equal to $(n+1)(h+1)-1$ where $n$ 
the arity of $x$.

{\bd Proof} The substitution $\sigma$ is a solution to the problem 
$\Phi(\Psi, \sigma)$, thus, by lemma 3, this problem has a solution $\sigma'$
such that for every instantiable variable $x$, $\sigma' x$ has a depth
less than or equal to $(n+1)(h+1)-1$. 
This solution $\sigma'$ is a solution to the problem $\Psi$.
\end{Lem}

\beginex{Remark} This method, in which an interpolation problem
$\Phi(\Psi,\sigma)$ is constructed from a pair $<\Psi,\sigma>$ where
$\Psi$ is an arbitrary problem and 
$\sigma$ a solution to $\Psi$, can be compared to the one
used in the completeness proof of \cite{SnyGal} in which a problem in
solved form is constructed from such a pair.
\endex

\section{A Decision Procedure}

\beginex{Theorem} Third Order Matching is Decidable

{\bd Proof}
A decision procedure is obtained by considering the problem $\Phi$ and
enumerating all the ground substitutions such that the term
substituted for $x$
has a depth less than or equal to $(n+1)(h+1)-1$, where $h$ maximum depth of $b$
for $a = b \in \Phi$ and $n$ is the arity of $x$. If one of these 
substitutions is a solution then success else failure. This decision 
procedure is obviously sound. By lemma 4, it is complete.
\endex

\beginex{Remark} A more efficient decision algorithm is obtained by
enumerating the nodes of the tree obtained by pruning Huet's search
tree \cite{Huet75} \cite{Huet76} at each node corresponding to a
substitution whose depth is larger than $(n+1)(h+1)-1$. This tree is
obviously finite and thus this algorithm terminates. It is obviously
sound. By lemma 4, it is complete.
\endex

\beginex{Remark}
This result can be used to design an algorithm which enumerates a
complete set of solutions to a third order matching problem and either
terminates
if the problem has a finite complete set of solutions or keeps
enumerating solutions forever if it the problem admits no such set.
Such an algorithm is got by enumerating the nodes of the tree obtained 
by pruning Huet's search tree \cite{Huet75} \cite{Huet76} at each node
labeled with a problem that has no solution (by the theorem above, it is
decidable if such a problem has a solution or not). 
Obviously, this algorithms still produces a complete set of solutions.

Let us show now that when a matching problem has a finite complete set
of solutions then this algorithm terminates.
Recall that a set of substitutions is called {\it minimal} if no substitution
of this set is an instance of another and that Huet's algorithm
applied to a matching problem produces a minimal complete set of
solutions \cite{Huet76}. It is routine to verify that if a a problem
has a finite complete set of solutions then any minimal complete set
of solutions is also finite. So, if a problem has a finite complete
set of solutions then Huet's tree for this problem has a
finite number of success nodes and thus a finite number of nodes
labeled with a problem that has a solution. The pruned tree is therefore
finite and the algorithm obtained by enumerating its nodes terminates.
\endex

\beginex{Remark}
This decidability result can be compared with the decidability of the 
equations of the form $P(x_{1}, ..., x_{n}) = b$ where $P$ is a polynomial 
whose coefficients are natural numbers and $b$ is a natural number. 

If this equation has a solution $<a_{1}, ..., a_{n}>$ then it has a solution 
$<a'_{1}, ..., a'_{n}>$ such that $a'_{1} \leq b$.
Indeed either $Q(X) = P(X,a_{2}, ..., a_{n})$ is not a constant polynomial and
for all $n$, $Q(n) \geq n$, so $a_{1} \leq b$, or the polynomial $Q$ is 
identically equal to $b$ and $<0, a_{2}, ..., a_{n}>$ is also a solution. So 
a simple induction on $n$ proves that if the equation has a solution then it 
also has a solution in $\{0, ..., b\}^{n}$ and an enumeration 
of this set gives a decision procedure.

\endex

\section*{Conclusion: Towards Higher Order Matching}

The proof given here is based on the fact that if $t$ is a third order term
then when we reduce the term $(t~c_{1}~...~c_{n})$,
{\it in the general case}, we get a term deeper than $t$ (or, at least, if it 
is not, the depth loss can bounded).
This gives a bound (in terms of the depth of $b$) on the depth of the 
solutions of the
equation $(x~c_{1}~...~c_{n}) = b$. 
In {\it the particular cases} in which the depth loss is greater than the 
bound, some part of the term $t$ is superfluous and that we can construct a 
smaller term $t'$ such that $(t'~c_{1}~...~c_{n}) = (t~c_{1}~...~c_{n})$. 

Generalizing this property of reduction to the full $\lambda$-calculus would
give the decidability of higher order matching. 
To get the normal form of the term $(t~c_{1}~...~c_{n})$ we have followed the
strategy hinted by the weak normalization theorem and reduced first 
all the second order redexes, then all the first order redexes. 
So a generalization of this proof to higher order should require an induction
on the maximal order of a redex. 
In the proof for the third order case, we quickly get the normal form of
the term $(t~c_{1}~...~c_{n})$ and we do not need to define the depth of a 
non-normal term. It seems that the generalization of this result to higher 
order requires such a definition.

\section*{Acknowledgments}

The author would like to thank G\'{e}rard Huet, Richard Statman 
and Gopalan Nadathur for many very helpful discussions on this problem and
remarks on previous drafts of this paper. This research was partly supported by
ESPRIT Basic Research Action ``Logical Frameworks''.


\begin{thebibliography}{99.}

\bibitem{Barendregt81}
H. Barendregt, The Lambda Calculus, its Syntax and Semantics, {\it North 
Holland}, 1981, 1984.

\bibitem{deBruijn72}
N.G. de Bruijn,
Lambda Calculus Notation with Nameless Dummies, a Tool for Automatic Formula
Manipulation, with Application to the Church-Rosser Theorem,
{\it Indagationes Mathematicae}, 34, 5, 1972, pp. 381-392.

\bibitem{Gorn}
S. Gorn,
Explicit Definitions and Linguistic Dominoes,
{\it University of Toronto}, 1967.

\bibitem{HinSel}
J.R. Hindley, J.P. Seldin,
Introduction to Combinators and $\lambda$-Calculus, {\it Cambridge University
Press}, 1986.

\bibitem{Huet75}
G. Huet,
A Unification Algorithm for Typed $\lambda$-calculus,
{\it Theoretical Computer Science}, 1, 1975, pp. 27-57.

\bibitem{Huet76}
G. Huet,
R\'{e}solution d'\'{E}quations dans les Langages d'Ordre 1,2, ...,
$\omega$,
{\it Th\`{e}se de Doctorat d'\'{E}tat}, Universit\'{e} de Paris VII, 1976.

\bibitem{HueLan}
G. Huet, B. Lang,
Proving and Applying Program Transformations Expressed with Second Order
Patterns,
{\it Acta Informatica}, 11, 1978, pp. 31-55.

\bibitem{Miller91}
D. A. Miller,
Unification Under a Mixed Prefix, {\it Journal of Symbolic
Computation}, 14, 1992, pp. 321-358.

\bibitem{SnyGal}
W. Snyder, J. Gallier, Higher-Order Unification Revisited: Complete
Sets of Transformations, {\it Journal of Symbolic Computation}, 8,
1989, pp. 101-140.

\bibitem{Statman}
R. Statman,
Completeness, Invariance and $\lambda$-definability,
{\it Journal of Symbolic Logic}, 47, 1, 1982, pp. 17-26.

\bibitem{Wolfram}
D.A. Wolfram,
The Clausal Theory of Types,
{\it PhD Thesis}, University of Cambridge, 1989.

\bibitem{Zaionc}
M. Zaionc,
The Set of Unifiers in Typed $\lambda$-Calculus as Regular Expression, 
{\it Proceedings of Rewriting Techniques and Applications}, 
Lecture Notes in Computer Science 202, Springer-Verlag, 1985, pp. 430-440.

\end{thebibliography}
\end{document}